\documentclass[journal,twoside,web]{ieeecolor}
\usepackage{jsen}
\usepackage{cite}
\usepackage{amsmath,amssymb,amsfonts}
\usepackage{algorithmic}
\usepackage{graphicx}
\usepackage{textcomp}
\usepackage{wrapfig}
\usepackage{bbm}

\def\BibTeX{{\rm B\kern-.05em{\sc i\kern-.025em b}\kern-.08em
    T\kern-.1667em\lower.7ex\hbox{E}\kern-.125emX}}
\definecolor{abstractbg}{rgb}{0.89804,0.94510,0.83137}
\begin{document}
\title{Real-time division-of-focal-plane polarization imaging system with progressive networks}
\author{Rongyuan Wu, Yongqiang Zhao, \IEEEmembership{Member, IEEE}, Ning Li, and Seong G.Kong, \IEEEmembership{Senior Member, IEEE}
\thanks{This work was supported by the National Natural Science Foundation
	of China (NSFC) under Grant 61771391, Key R \& D plan of Shaanxi
	Province 2020ZDLGY07-11, the Science, Technology and Innovation
	Commission of Shenzhen Municipality under Grants
	JCYJ20170815162956949 and JCYJ20180306171146740 and the
	National Research Foundation of Korea (2016R1D1A1B01008522).}
\thanks{R. Wu, Y. Zhao, N. Li are with the School of Automation, Northwestern Polytechnical University, Xi’an 710072, China (e-mail: rongyuanwu@mail.nwpu.edu.cn; zhaoyq@nwpu.edu.cn; ln\underline{~}neo@mail.nwpu.edu.cn). }
\thanks{S. G. Kong is with the Department of Computer Engineering, Sejong
	University, Seoul 05006, Korea (e-mail: skong@sejong.edu).}
}

\IEEEtitleabstractindextext{%
\fcolorbox{abstractbg}{abstractbg}{%
\begin{minipage}{\textwidth}%
\begin{abstract}
Division-of-focal-plane (DoFP) polarization imaging technical recently has been applied in many fields. However, the images captured by such sensors cannot be used directly because they suffer from instantaneous field-of-view errors and low resolution problem. This paper builds a fast DoFP demosaicing system with proposed progressive polarization demosaicing convolutional neural network (PPDN), which is specifically designed for edge-side GPU devices like Navidia Jetson TX2.
The proposed network consists of two parts: reconstruction stage and refining stage. The former recovers four polarization channels from a single DoFP image. The latter fine-tune the four channels to obtain more accurate polarization information. PPDN can be implemented in another version: PPDN-L (large), for the platforms of high computing resources. Experiments show that PPDN can compete with the best existing methods with fewer parameters and faster inference speed and meet the real-time demands of imaging system.
\end{abstract}

\begin{IEEEkeywords}
Division-of-focal-plane polarization imaging system, polarization demosaicing, progressive convolutional neural network, efficiency.
\end{IEEEkeywords}
\end{minipage}}}

\maketitle

\section{Introduction}
\label{sec:introduction}
\IEEEPARstart{P}{olarization} imaging has been successfully applied to the applications such as road detection \cite{li2020full,li2021illumination}, histopathology diagnosis \cite{zhao2020detecting}, detection of defects \cite{li2021multisensor}, reflection removal \cite{li2018removal}, and image dehazing \cite{shen2018iterative} thanks to its perceptual capabilities over traditional imaging. To get polarization signals, it is essential to obtain intensity images of different orientations. Division of time (DoT) \cite{harnett2002liquid}, division of amplitude (DoAM) \cite{farlow2002imaging}, division of aperture (DoAP) \cite{tyo2006hybrid}, and division-of-focal-plane (DoFP) polarimeters \cite{perkins2010signal} are the most common types of polarization imaging sensors now available. The DoT has the apparent problem of requiring a static picture. In other words, interframe motion will result in erroneous polarization information. The DoAM system is difficult to align, and the incident intensity is reduced using beam splitters. It is also difficult to align the DoAP system, and it is susceptible to polarization-dependent aberration effects. DoFP polarimeter has been a mainstream polarization imaging technology since it can instantaneously capture dynamic polarization information. DoFP polarimeters capture an image with multiple polarization angles using a micro-polarizer array.  Figure. \ref{fig:sys} shows a micro-polarizer array in four orientations (0°, 45°, 90°, and 135°) in a period of $2\times2$ from the Sony IMX250MZR CMOS. To obtain the polarization signals from a DoFP image, a demosaicing process is needed. Polarization demosaicing (PDM) is an essential imaging pipeline to reconstruct four polarization orientation images from a DoFP image. Unlike color image demosaicing, polarization demosaicing algorithms are evaluated not only on the four reconstructed polarization images, but more crucial, on the three important parameters: intensity ($S_{0}$), $DoLP$, and $AoLP$, which are calculated as \cite{goldstein2017polarized}:

\begin{subequations}
	\begin{align}
	S_{0} &= 0.5*(I_{0} + I_{45} + I_{90} + I_{135})  \text{,}
	\label{1a}  \\
	S_{1} &= I_{0} - I_{90} \text{,}
	\label{1b}  \\
	S_{2} &= I_{45} - I_{135}  \text{,}
	\label{1c}  \\
	DoLP &= \sqrt{S_{1}^{2} + S_{2}^{2}} / S_{0}  \text{,}
	\label{1d}  \\
	AoLP &= 0.5*\arctan(S_{2} / S_{1})  \text{,}
	\label{1e}  
	\end{align}
\end{subequations}

where $I_{0}$, $I_{45}$, $I_{90}$, $I_{135}$ are captured intensity images at four polarization orientations (0°, 45°, 90°, and 135°), respectively. Stokes parameter $S_{0}$ denotes the total intensity of the light. $S_{1}$ and $S_{2}$ describe the linear polarization states. $DoLP$ indicates the strength of polarization and $AoLP$ denotes the direction of polarization.

Many color \cite{bayer1976color,mairal2009non,kiku2013residual,tan2017color,cui2018color} and multispectral image demosaicing methods \cite{monno2012multispectral,mizutani2014multispectral,mihoubi2017multispectral,shinoda2018deep,feng2021mosaic} have been suggested, but none of them are directly applicable to polarization image demosaicing. The CFA is usually in Bayer pattern, with the three color channels of red, green, and blue organized in a 1:2:1 ratio. MSFA are typically 3$\times$3 or 4$\times$4 in size which is completely different from the 2$\times$2 design of PFA. Polarization image demosaicing is fundamentally different from the previous two in terms of both physical meaning and filter array design, as it focuses not only on the quality of the reconstructed single polarization channel image, but also on the quality of the polarization signal calculated from the four recovered polarization images.
Demosaicing techniques based on interpolation \cite{zhang2016image,li2019demosaicking,wu2021polarization} have been proposed to solve the PDM problem, however, the reconstructed images often lack high-frequency information. This may cause artifacts in intensity images and the reconstructed polarization information can be less accurate. Researchers have also investigated deep-learning for polarization demosaicing. Zhang et al. \cite{zhang2018learning} propose a network with skip connections called PDCNN to learn an end-to-end mapping between the coarse interpolation results and full-resolution polarization images. PDCNN takes bicubic-interpolated results as the input, which introduces interpolation error to the network with inaccurate reconstructed results. Zeng et al. \cite{zeng2019end} propose a four-layer convolutional network called ForkNet. It learns an end-to-end mapping from DoFP images to $S_{0}$, $DoLP$, and $AoLP$. ForkNet uses only one convolution layer on the shared feature map to distinguish them, which can not recover $AoLP$ well. In addition to considering algorithm performance, the actual deployment is another important factor for real-time polarization imaging systems.

This paper builts a real-time DoFP imaging system with a progressive polarization demosaicing network (PPDN) using a coarse-to-fine scheme. PPDN takes a DoFP image as the input instead of an interpolation result, which eludes interpolation errors and thus facilitate feature extraction of the network. Although Zeng et al. \cite{zeng2019end} claim that direct reconstruction of $S_{0}$, $DoLP$, and $AoLP$ can avoid cumulative errors compared to reconstructing four polarization images, PPDN can still minimize the errors. PPDN provides a two-stage approach to reconstruct the intensity information as well as the polarization information. Furthermore, we provide a more physically relevant loss function for $AoLP$ reconstruction, which is utilized to compute the reconstruction loss precisely around 0° $AoLP$. For real-time polarization imaging systems, practical deployment is an important factor in addition to algorithm performance. The actual deployment must take into account influencing factors like model size, memory usage while the model is running, and model inference speed. In terms of computational complexity, number of model parameters, and actual inference speed, the proposed PPDN can meet the deployment requirements well while maintaining high accuracy.

The contributions of this paper are:

(1) Built a real-time DoFP polarization imaging system, which can surve as a base for subsequent tasks such as object detection and segmentation.

(2) Proposed a progressive polarization demosaicing network with a physically meaningful AoLP loss function, which recovers the intensity information and polarization information of the DoFP image in a phased manner.  

(3) Built a color-polarization dataset to validate the proposed algorithm, which contains a large number of indoor and outdoor scenes to simulate different polarization imaging conditions.

(4) Verified real-time performance and stability of the
proposed algorithm on the system we built. 

\section{Real-time Polariztion Imaging System}

\subsection{DoFP Image Acquisition and Demosaicing System}
\begin{figure}[h]
	\begin{center}
		\includegraphics[width=3.0in]{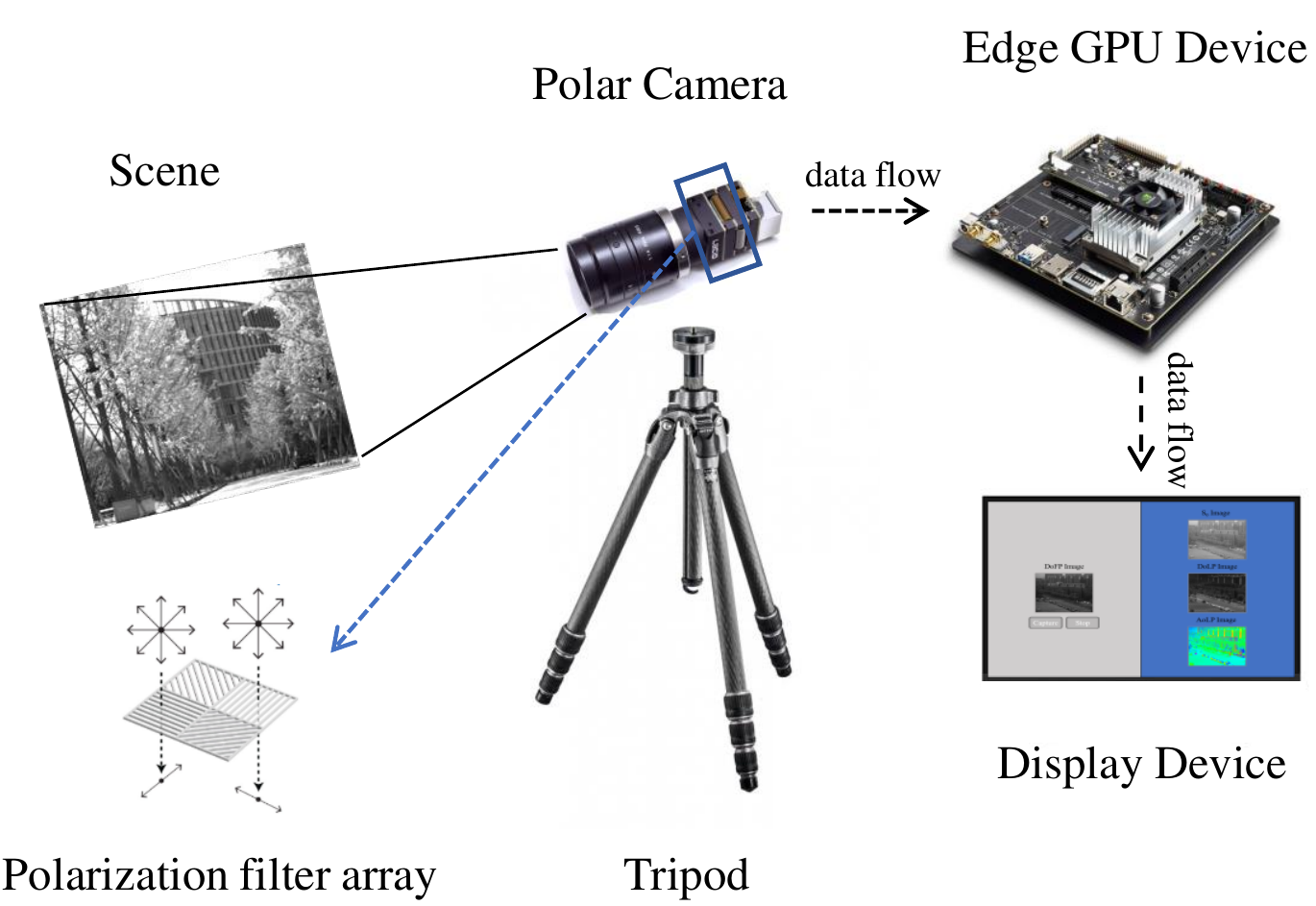}
	\end{center}
	\caption{Real-time DoFP Image Acquisition and Demosaicing System.}
	\label{fig:sys}
\end{figure}

Figure \ref{fig:sys} shows a real-time DoFP image acquisition and demosaicing system. To record the polarization information of the scene, a 2048$\times$2448 visible polarization camera (Sony PHX050S-PC) with four distinct directional micro polarizers (0°, 45°, 90°, and 135°) combined on a single chip is utilized. The Nvidia Jetson TX2 is used as an edge GPU device for picture capture control and network inference before transferring the reconstruction polarization parameters ($S_{0}$, $DoLP$, and $AoLP$) to the display device. The real-time DoFP polarization imaging system we built can not only accurately obtain the scene's polarization information, but also, more importantly, lay the groundwork for using the scene's polarization information for object tracking, pedestrian detection, and other tasks, and their models can be directly integrated into the system's already built system to improve its functions.
\subsection{Progressive Polariztion Demosaicing Network}
The goal of PPDN is to learn a mapping $F_{\theta}$ in a coarse-to-fine way between DoFP images $y\in \textbf{R}^{H\times W}$ and four polarization images $X\in \textbf{R}^{H\times W\times 4}$ as:
\begin{align}
\begin{split}
X = F_{\theta}(y) \text{,}
\end{split}
\end{align}
where $\theta$ are the parameters of the network.

\begin{figure}[t]
	\begin{center}
		\includegraphics[width=2.8in]{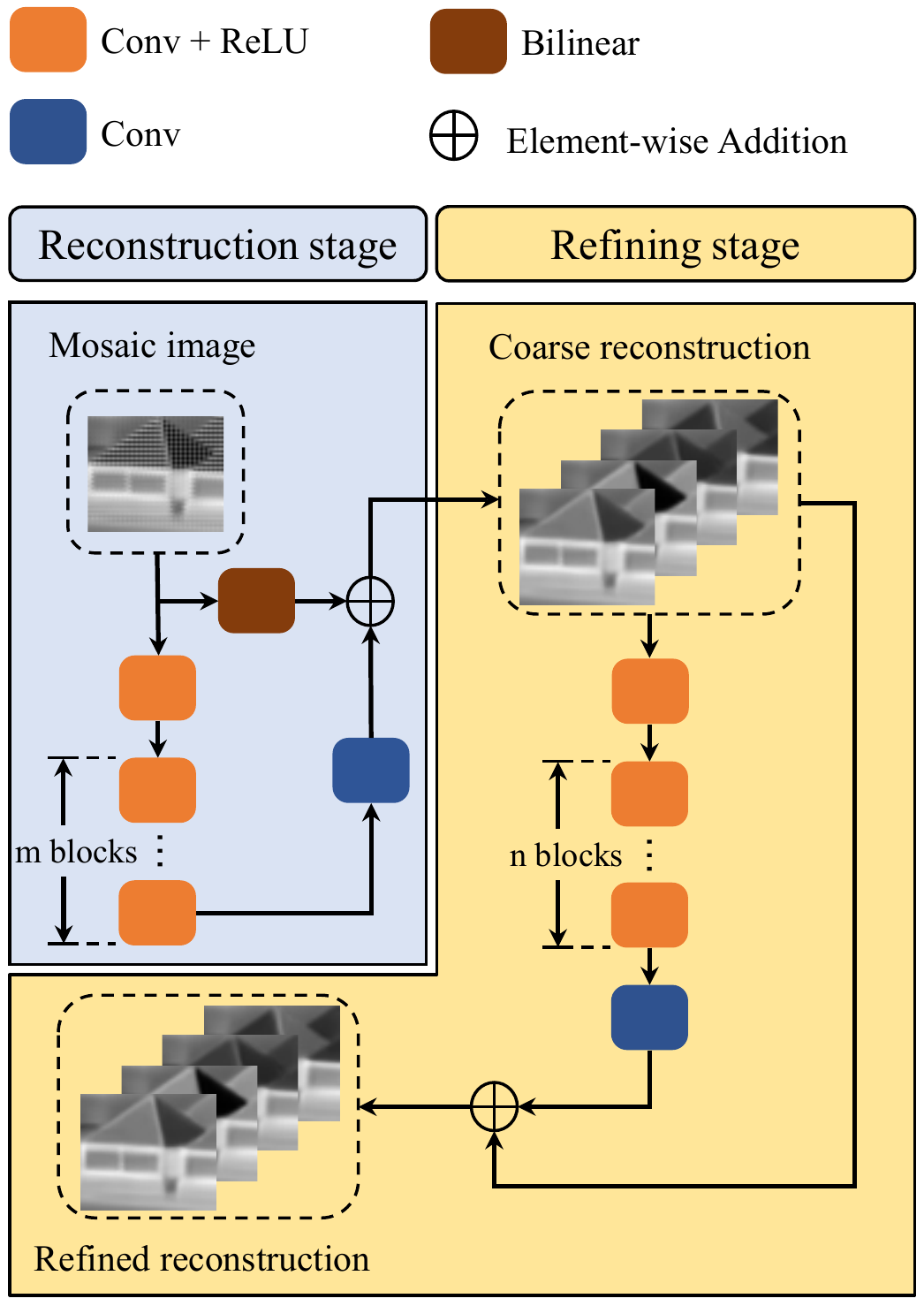}
	\end{center}
	\caption{The overview of the proposed PPDN. PPDN is a two-stage network framework that reconstructs four full-resolution polarization images in a coarse-to-fine way.}
	\label{fig:outline}
\end{figure}

As illustrated in Fig. \ref{fig:outline}, at the reconstruction stage, the DoFP images are considered as the input. The ‘clean’ input without interpolation errors might help the network to perform feature extraction more accurately. Bilinear branch (BB) is completed with a fixed convolution kernel. The trainable branch (TB) is used for extracting contextual features from input DoFP images and learning the residual between ground-truth and bilinear interpolation results. In the refining stage, a simple residual learning strategy is applied to further fine-tune the coarse reconstruction to improve the reconstructed polarization information.
For the reconstruction stage, we extract the shallow features $F_{s}$ from the input DoFP image $y$ in the trainable branch by 
\begin{align}
\begin{split}
F_{s} = \delta(C(y)) \text{,}
\end{split}
\end{align}
where $C$ denotes the convolution operation with $3\times3$ filters. $\delta$ represents the rectified linear unit (ReLU) activation function. Then $F_{s}$ is sent to multiple basic blocks in series for finer feature extraction.
\begin{align}
\begin{split}
F_{m} = B_{m}(B_{m-1}(\cdots B_{1}(F_{s})\cdots)) \text{,}
\end{split}
\end{align}
where $B_{m}$ denotes the $m$-$th$ basic block, which is composed of the convolution layer with $3\times3$ filters and ReLU. Another branch of the reconstruction stage produces bilinear interpolation results $X_{bi}$ with a fixed convolution kernel as \cite{wu2021polarization}. The coarse reconstruction result $X_{cr}\in\textbf{R}^{H\times W\times 4}$ is obtained by adding up the $X_{bi}\in\textbf{R}^{H\times W\times 4}$ and a residual prediction produced by $F_{m}$ as:
\begin{align}
\begin{split}
X_{cr} = X_{bi} + C(F_{m}) \text{,}
\end{split}
\end{align}

For the refining stage (RS), we improve $X_{cr}$ by repairing the local details in the form of residual:
\begin{align}
\begin{split}
X_{rr} = X_{cr} + C(B_{n+1}(\cdots B_{1}(X_{cr})\cdots)) \text{,}
\end{split}
\end{align}
The filter numbers of each convolution layer are set to $k$, which can be changed to adjust the representational power of the model for the platforms of high and low computing resources.

\subsubsection{Loss Function}
The reconstruction stage and the refining stage are optimized jointly with the loss function defined as:
\begin{align}
\begin{split}
L = & w_{1}(L_{m}(X_{cr}, X) + L_{m}(S_{0,cr}, S_{0})) \\
+ & w_{2}(L_{m}(X_{rr}, X) + L_{m}(S_{0,rr}, S_{0})) \\
+ & w_{3}(L_{m}(S_{1,rr}, S_{1}) + L_{m}(S_{2,rr}, S_{2})  \\ 
+ & L_{m}(DoLP_{rr}, DoLP)) \\
+ & w_{4}(L_{m,aolp}(AoLP_{rr}, AoLP)) \text{,}
\end{split}
\end{align}
where $S_{0}$, $S_{1}$, $S_{2}$, $DoLP$ and $AoLP$ are the ground-truth images. $S_{0,cr}$ is calculated by $X_{cr}$ from Eqs (1). $S_{0,rr}$, $S_{1,rr}$, $S_{2,rr}$, $DoLP_{rr}$ and $AoLP_{rr}$ are calculated by $X_{rr}$. $L_{m}$ is the mean absolute error (MAE) loss and $L_{m,aolp}$ is the MAE specially designed for $AoLP$, which is explained in detail below. $w_{1}$, $w_{2}$, $w_{3}$ and $w_{4}$ are the balancing parameters. The network does not put polarization limitations on the coarse recovery outcomes. Because it intends to focus on recovering intensity information. 

\begin{figure}[t]
	\begin{center}
		\includegraphics[width=2.8in]{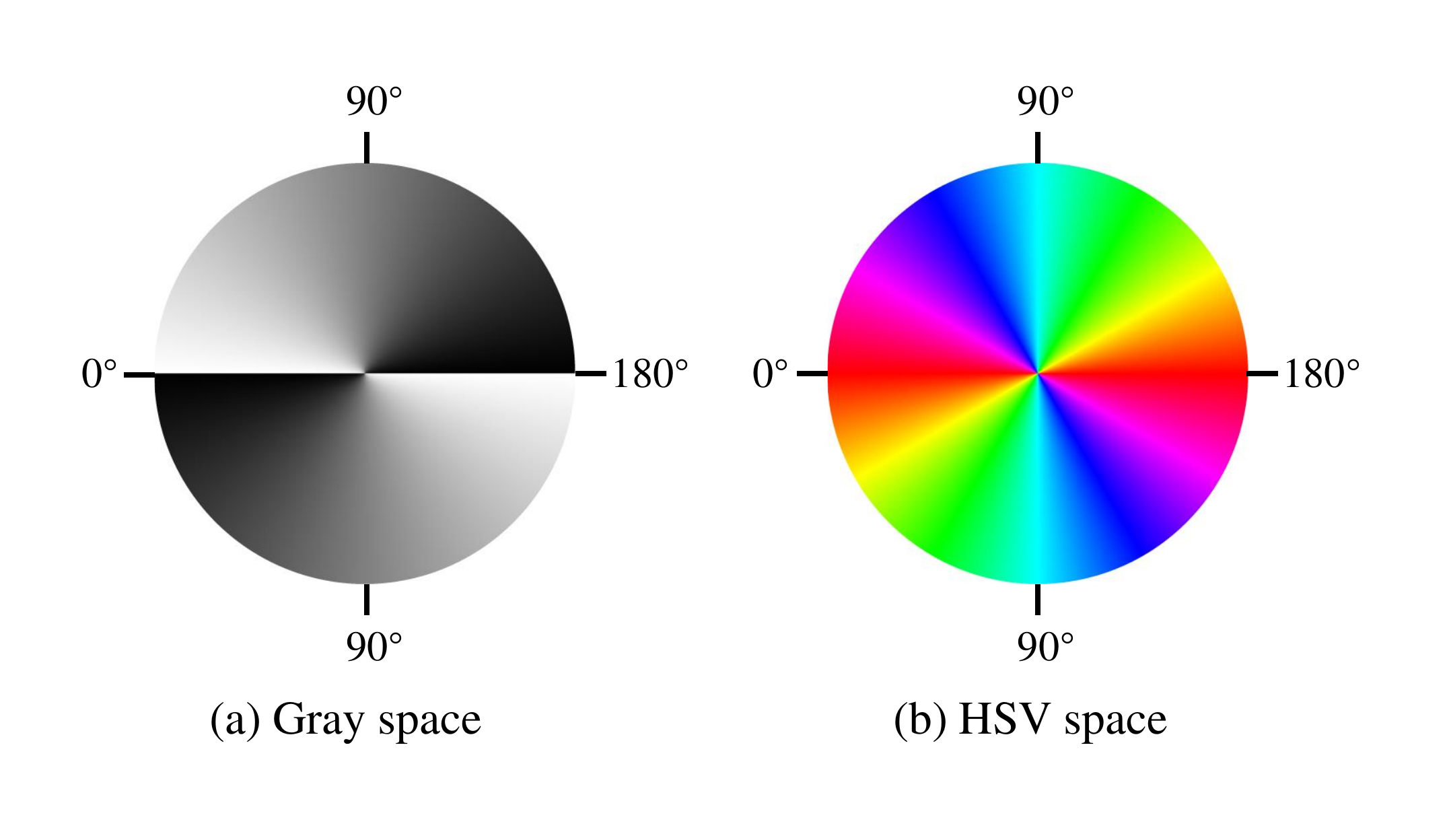}
	\end{center}
	\caption{Display difference of $AoLP$ between grayscale space and HSV space.}
	\label{fig:aolp}
\end{figure}

$AoLP$'s  range is 0°-180°, with 0° and 180° denoting the same polarization angle direction. However, the grayscale image maps it to 0-1 as Fig. \ref{fig:aolp} (a), and calculating the loss function of the polarization angle using the ordinary MAE can easily introduce a large error in the 0° direction (0 and 1 indicate the same physical meaning, but the error is the largest at this point), so we borrow the rule of HSV spatial display (Fig. \ref{fig:aolp} (b)) and calculate the loss function of the polarization angle using a closer distance on the circle.  $L_{m,aolp}$ is defined as:

\begin{align}
\begin{split}
& L_{m,aolp}(Ref, input) = \\
& min(||Ref - input||_{1}, 1 - ||Ref - input||_{1}) \text{,}
\end{split}
\end{align}
where $Ref$ is the ground-truth image and $input$ is the reconstructed $AoLP$ image.

\subsubsection{Training Details}
For training, we use $64\times64$ patches cropped from synthetic DoFP images as the input and augment the training data with random flips (vertical and horizontal) and rotations (90°, 180°, and 270°). We train our model with an ADAM optimizer \cite{kingma2014adam} by setting $\beta_{1}$ = 0.9, $\beta_{2}$ = 0.999. The mini-batch size is set to 16. The initial learning rate is set to $3\times 10^{-4}$ and decreases to 0.1 times every 48K iterations. And the number of total iterations is 96K. The balancing parameters $w_{1}$, $w_{2}$, $w_{3}$ and $w_{4}$ in Eq (7) are empirically set to 0.25, 0.5, 1 and 0.1, respectively. We implement our model with Pytorch and run experiments with an RTX 2080 Ti GPU. 	

\section{Color and polarization image dataset}
\begin{figure}[h]
	\begin{center}
		\includegraphics[width=3.5in]{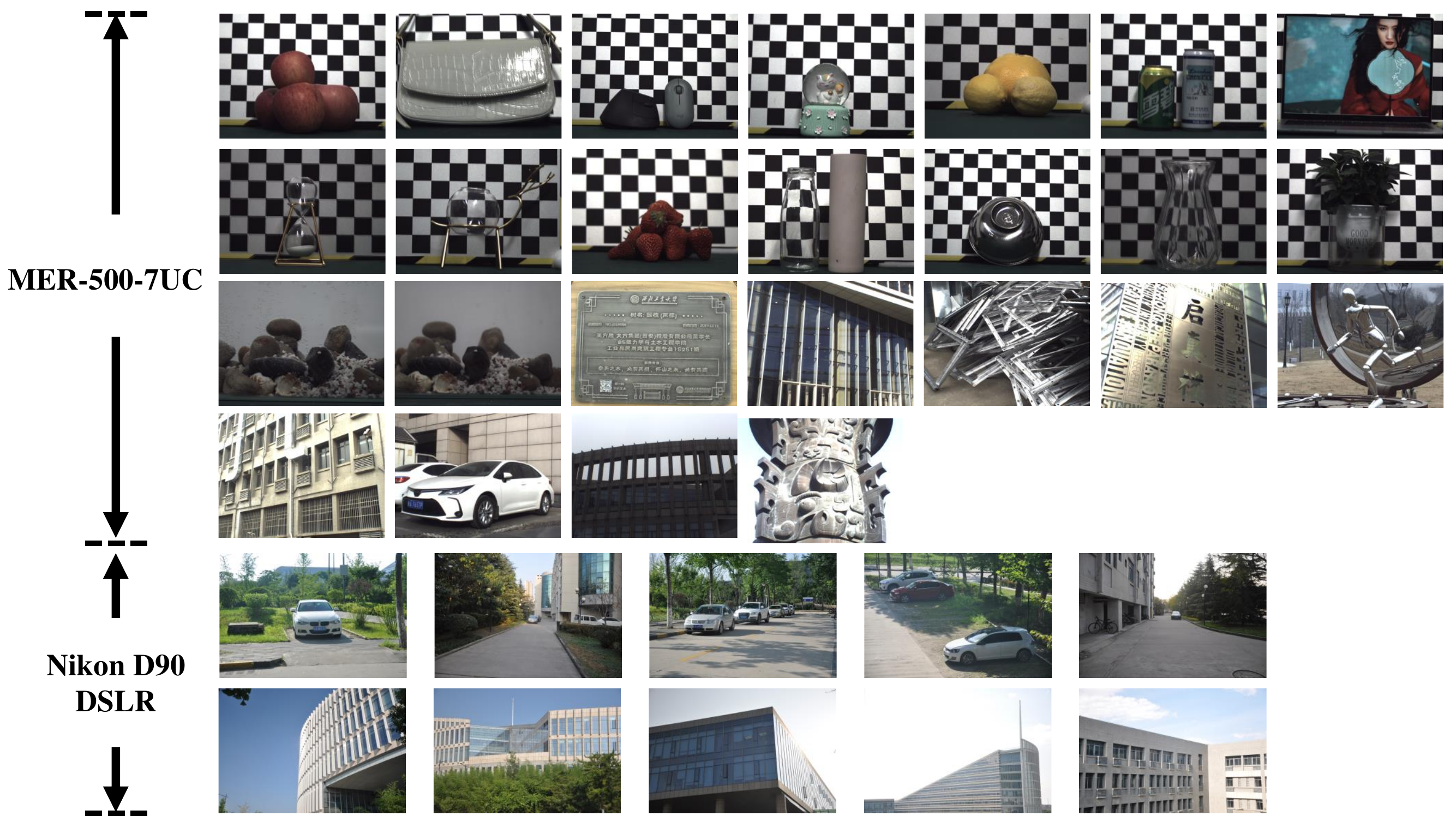}
	\end{center}
	\caption{Gallery of our polarization image dataset. Only the intensity components are displayed in this diagram.}
	\label{fig:dataset}
\end{figure}
To the best of our knowledge, most of the scenes in the existing polarization database \cite{lapray2018database, qiu2019polarization, morimatsu2020monochrome} are inside, resulting in a significant imbalance in outside situations. Outdoor situations indicate a more complicated lighting environment, which is crucial for assessing the robustness of polarization demosaicing methods. Furthermore, the previous database was built using the same industrial cameras throughout the acquisition process, resulting in very consistent data distribution, which is negative for network training because the generalization performance of a well-trained network based on a specific data distribution is limited. Indoor scenes are filmed with calibration plates rather than walls as backdrops to improve the scene's polarization properties. As a result, we take 37 sets of the scenarios, 16 of which are indoor scenes  and 21 of which are outdoor scenes. We attempted to include as many diverse materials, shapes, and lighting situations as possible. Two sets of scenes imitating underwater polarization imaging are also included in the database (one group has bubbles while the other does not). As mentioned in \cite{lapray2018database}, all data are meticulously aligned using the toolbox \cite{IAT2013}. We use a MER-500-7UC color camera with 1944$\times$2592 pixel size to collect the first 27 sets of the scenarios and a Nikon D90 DSLR camera with 2750$\times$4190 pixel size for another 10 sets of the scenarios. Figure \ref{fig:dataset} shows a gallery of our polarization image datasets.

The linear polarizer is placed in front of the camera. By spinning the polarizer in different orientations (0°, 45°, 90°, and 135°), we collect four sets of raw pictures for each scenario. To reduce noise, 200 frames per orientation are recorded for average. In the averaged picture, we additionally conduct a 2 $\times$ 2 pixel binning to improve the signal-to-noise ratio as \cite{qiu2019polarization}.

\section{Experiment Results}

\subsection{Accuracy Comparison on Simulation Data}
PPDN is proposed for monochrome DoFP images demosaicing, so we use the green channel as the ground truth. Synthetic DoFP images are generated by down-sampling the four captured polarization images as the pattern in Fig. \ref{fig:sys}. We randomly select 11 indoor scene images and 6 outdoor scene images taken by MER-500-7UC color camera, and 6 outdoor scene images taken by Nikon D90 DSLR camera as the training set, and the rest as the test set. We design two different size models which are named PPDN and PPDN-L by adjusting the filter numbers $k$ of each convolution layer, the number of basic blocks $m$ in the reconstruction stage and the number of basic blocks $n$ in the refining stage. For PPDN, we set $m$=$n$=1 and $k$=32 to verify the function of the framework at few computational costs. For PPDN-L, we set $m$=9, $n$=3 and $k$=64 to explore the power of the proposed framework. We compare the proposed method with several mainstream polarization demosaicing methods such as NP \cite{li2019demosaicking}, ForkNet \cite{zeng2019end}, and PDCNN \cite{zhang2018learning}. NP is the traditional interpolation method, PDCNN and Forknet are CNN-based methods. The peak signal-to-noise ratio (PSNR) is adopted as the quantitative criterion to evaluate the performance of different works. When calculating the PSNR value of the reconstruction $AoLP$, we calculate the mean squared error (MSE) in a HSV space similar with Eq (8). Table \ref{tab:evaluation} shows that the proposed PPDN-L outperforms other state-of-the-art (SOTA) methods by a large margin and maintains acceptable computational and parametric quantities. The performance of the proposed PPDN is similar to that of PDCNN. The polarization signals, in particular, are recreated more precisely. Although ForkNet is a lightweight model, its recovery ability seems to be limited, especially in the $AoLP$. 
\begin{table}[h]
	\centering
	\caption{\bf PSNR (dB) comparison of different methods on the test set}
	\begin{tabular}{cccccc}
		\hline
		& NP & ForkNet & PDCNN & PPDN & PPDN-L\\
		\hline
		$I_{0}$   	& 40.58 & --- 	& 43.69 & 43.33 & \textbf{44.72} \\
		$I_{45}$  	& 42.17 & --- 	& 44.40 & 44.16 & \textbf{45.67} \\
		$I_{90}$  	& 41.69 & --- 	& 45.24 & 44.64 & \textbf{46.38} \\
		$I_{135}$   & 42.47 & ---   & 45.34 & 45.08 & \textbf{46.45} \\
		$S_{0}$     & 45.60 & 45.85 & 48.70 & 47.98 & \textbf{49.76} \\
		$DoLP$      & 30.72 & 34.53 & 35.42 & 35.70 & \textbf{37.36} \\
		$AoLP$      & 28.35 & 13.82 & 31.01 & 31.03 & \textbf{32.52} \\
		\hline
	\end{tabular}
	\label{tab:evaluation}
\end{table}

\subsection{Real DoFP Image Demosaicing Comparison}
Using the built-in real-time DoFP image demosaicing system, we evaluate the performance of several models. Due to hardware limitations, both PDCNN and PPDN-L suffer from inadequate memory problems during model inference when dealing with 2048$\times$2448 image resolution input. To solve it, we split the pictures into four parts on average, input them into the network one by one, and then stitch the results together. As a result, utilizing a 1024$\times$1224 image resolution as input, their claimed findings in terms of computation and inference time consumption are increased by four. The stated issue does not exist in PPDN or ForkNet. The proposed PPDN has the least amount of computation, the lowest parameters and the fastest inference speed, as indicated in the Table \ref{tab:eva_real}. Moreover, PPDN-L is more lightweight than PDCNN and satisfies the criterion of real-time inference (25 FPS). 
\begin{table}[h]
	\centering
	\caption{\bf Computational complexity comparison, network parameters comparison and inference rate comparison of different methods on the real-time system}
	\begin{tabular}{cccccc}
		\hline
		& ForkNet & PDCNN & PPDN & PPDN-L\\
		\hline
		$Gflops$    & 302 & 2846 & \textbf{112} & 2260 \\
		$Param(K)$  & 60.3 	& 567.6 & \textbf{22.4} & 450.8 \\
		$FPS$       & 375 	& 21 & \textbf{380} & 34 \\
		\hline
	\end{tabular}
	\label{tab:eva_real}
\end{table}

\begin{figure}[t]
	\begin{center}
		\includegraphics[width=3.5in]{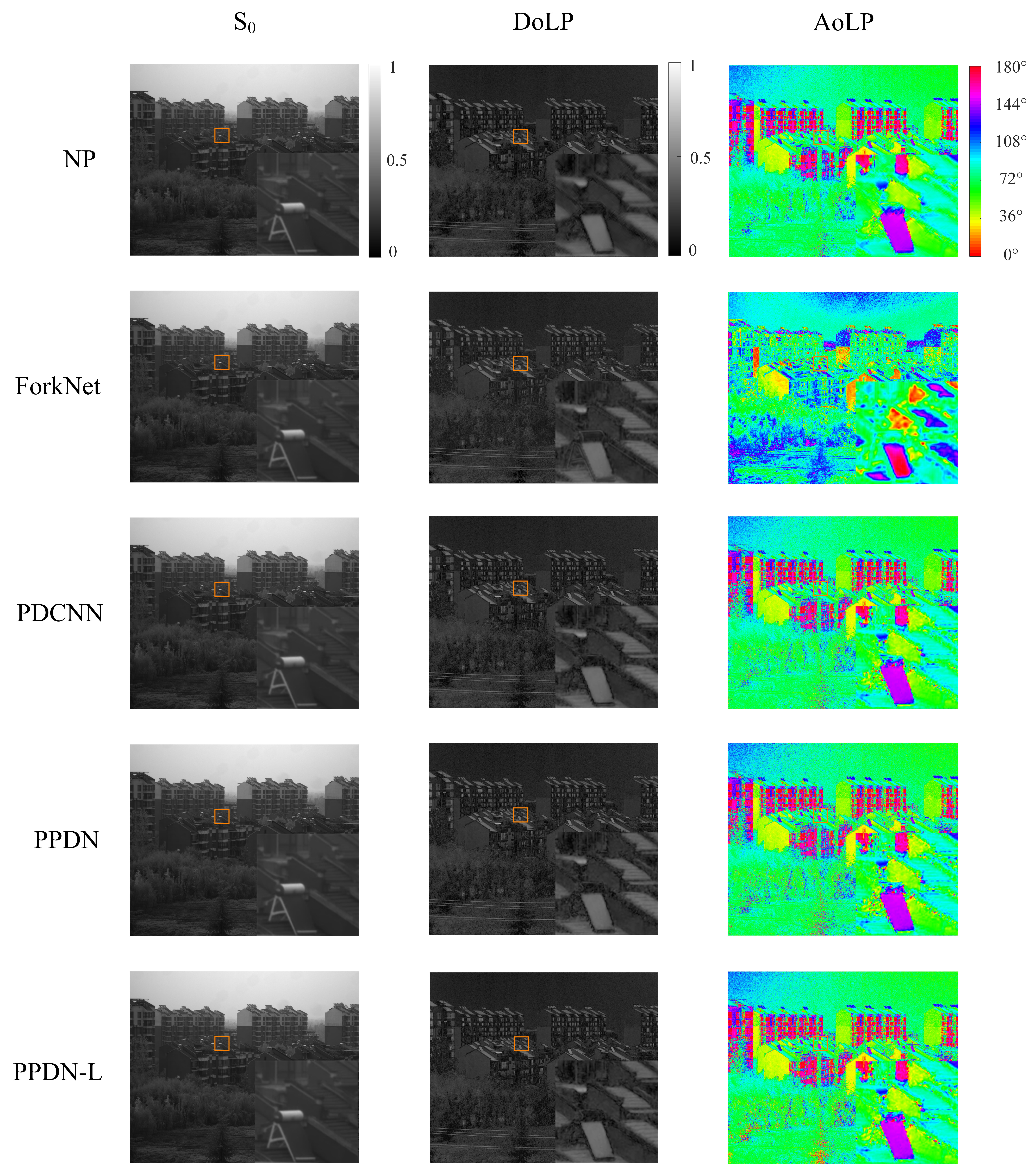}
	\end{center}
	\caption{Visual comparison of different demsoaicing methods for the real DoFP image captured by our system. Zoom in for more details.}
	\label{fig:vis}
\end{figure}

In Fig. \ref{fig:vis}, we provide recovery results of $S_{0}$, $DoLP$, $AoLP$ of each method. To reflect the physical properties of $AoLP$, we map it from grayscale space to HSV space for presentation. According to $DoLP$'s recovery findings, NP over-smoothes the support section of the solar water heater, resulting in deformed edges, ForkNet exhibites evident white pseudo-edges. In reconstructing the left edge of the solar panel, PPDN is less effective than PDCNN, but PDCNN appears white pseudo-edge at the lower edge, whereas PPDN does not. The PPDN-L recovery outcomes are among the best. Furthermore, ForkNet's $AoLP$ recovery findings reveal more significant distortion.

\subsection{Ablation Study}

For the network structure, we conduct an ablation experiment to explore the importance of each branch of the network. We remove BB, TB, and RS of the PPDN-L at a time and then retrain the network, keeping the training and testing process the same as before. Removing the TB means that the bilinear interpolation results are taken directly as input to the refining stage, and then we set $w_{1}$ to zero because the network does not need to optimize the results of bilinear interpolation. When removing the RS, we change the loss function to
\begin{align}
\begin{split}
L = & w_{2}(L_{m}(X_{cr}, X) + L_{m}(S_{0,cr}, S_{0})) \\
+ & w_{3}(L_{m}(S_{1,cr}, S_{1}) + L_{m}(S_{2,cr}, S_{2})  \\ 
+ & L_{m}(DoLP_{cr}, DoLP)) \\
+ & w_{4}(L_{m,aolp}(AoLP_{cr}, AoLP)) \text{,}
\end{split}
\end{align}
\begin{table}[t]
	\centering
	\caption{\bf PSNR (dB) comparison of net structure on the test set. \checkmark (-) indicates that the module is (not) added to the baseline network}
	\begin{tabular}{cccccc}
		\hline
		BB & TB & RS & $S_{0}$ & $DoLP$ & $AoLP$ \\ \hline
		-   & \checkmark  & \checkmark  & 45.06    & 31.72    & 16.74    \\
		\checkmark   & -  & \checkmark  & 49.21    & 36.95    & 32.20    \\
		\checkmark   & \checkmark  & -  & 49.61    & 33.39    & 29.36    \\
		\checkmark   & \checkmark  & \checkmark  & \textbf{49.76}  & \textbf{37.36}  & \textbf{32.52}  \\ \hline
	\end{tabular}
	\label{tab:aba_1}
\end{table}

\begin{table}[t]
	\centering
	\caption{\bf PSNR (dB) Comparison for the proposed $AoLP$ loss function.}
	\begin{tabular}{ccccc}
		\hline
		& $L_{m}$ & $L_{m, aolp}$ \\
		\hline
		
		$S_{0}$     & 47.85   & \textbf{47.98} \\
		$DoLP$      & 35.56   & \textbf{35.70} \\
		$AoLP$      & 30.83   & \textbf{31.03} \\
		
		\hline
	\end{tabular}
	\label{tab:aba_3}
\end{table}

For a fair comparison, we set $n$=12 when removing the TB and $m$=12 when removing the RS to keep their parameters at the same level. As shown in Table \ref{tab:aba_1}, train without BB will make it very difficult for the network to reconstruct the four polarization images from the DoFP images, Lacking TB in the reconstruction stage, the capability of the net seems limited, which also shows that directly using the interpolation results as network inputs would degrade the performance of the network. Cutting the RS can reconstruct as accurate $S_{0}$ as possible, however, the polarization information of the scene is severely lost even with the additional loss of $S_{1}$, $S_{2}$ and $DoLP$. This illustrates the importance of RS.

In order to verify the proposed $AoLP$ loss function $L_{m,aolp}$, we conduct comparison experiments on PPDN. One set of experiments uses the normal MAE loss function in calculating the aolp loss, and the other set uses the specially designed loss function. Table \ref{tab:aba_3} shows the outcomes of the experiments. The more physiologically relevant loss function aids not just $AoLP$ recovery but also $S_{0}$ and $DoLP$ gains.

\section{Conclusion}
On the basis of the Nvidia Jetson TX2, we develope a real-time DoFP polarization imaging system. To meet the demand for real-time imaging with restricted computing resources, we propose a lightweight demosaicing framework with polarization parameter reconstruction quality comparable to the best existing methods and higher inference efficiency. The improved version can meet the demand for real-time inference while providing a much better reconstruction effect than previous methods.

\section*{Acknowledgment}
 Rongyuan Wu thanks Master Jiaxiang Liu for the constructive discussions. We also thank Junchao Zhang for sharing the executable code.

\begin{IEEEbiography}[{\includegraphics[width=1in,height=1.25in,clip,keepaspectratio]{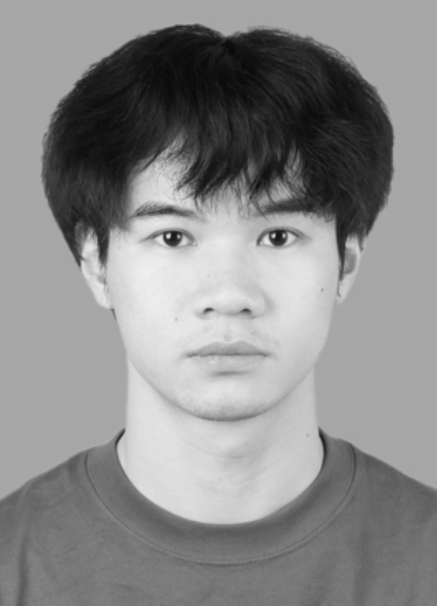}}]{Rongyuan Wu} received the B.S degree from the School of Automation, Northwestern Polytechnical University, Xi’an, in 2019, where he is currently pursuing the M.S. degree. 
	
	His current research interests include polarization imaging and computer vision, especially image demosaicking, denoising and super-resolution.
\end{IEEEbiography}

\begin{IEEEbiography}[{\includegraphics[width=1in,height=1.25in,clip,keepaspectratio]{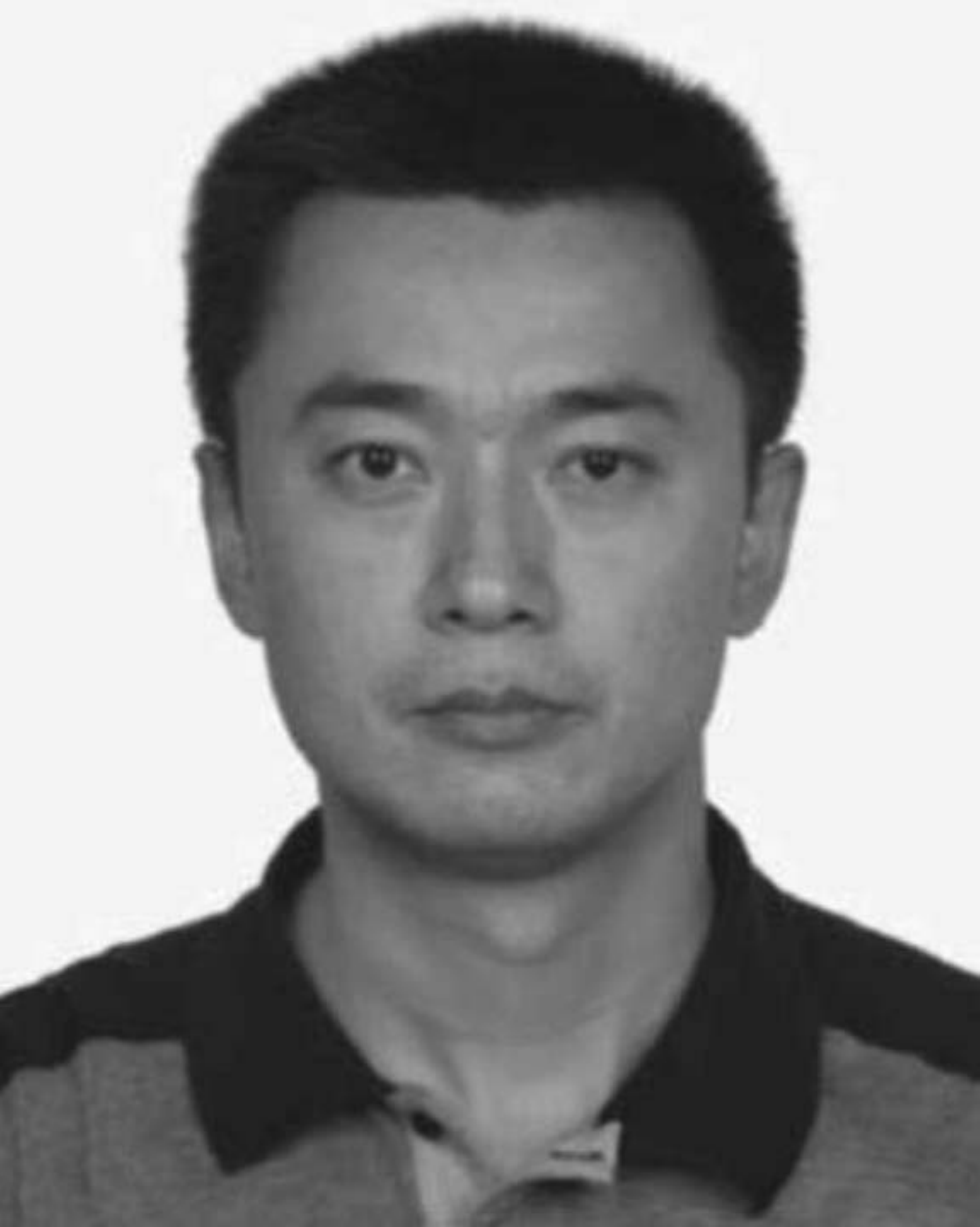}}]{Yongqiang Zhao} (Member, IEEE) received the B.S., M.S., and Ph.D. degrees in control science and engineering from Northwestern Polytechnic University, Xi’an, China, in 1998, 2001, and 2004, respectively.
	
	 From 2007 to 2009, he was a Post-Doctoral Researcher with McMaster University, Hamilton, ON, Canada, and Temple University, Philadelphia, PA, USA. He is a Professor with Northwestern Polytechnical University. His research interests include polarization vision, hyperspectral imaging, compressive sensing, and pattern recognition.
\end{IEEEbiography}

\begin{IEEEbiography}[{\includegraphics[width=1in,height=1.25in,clip,keepaspectratio]{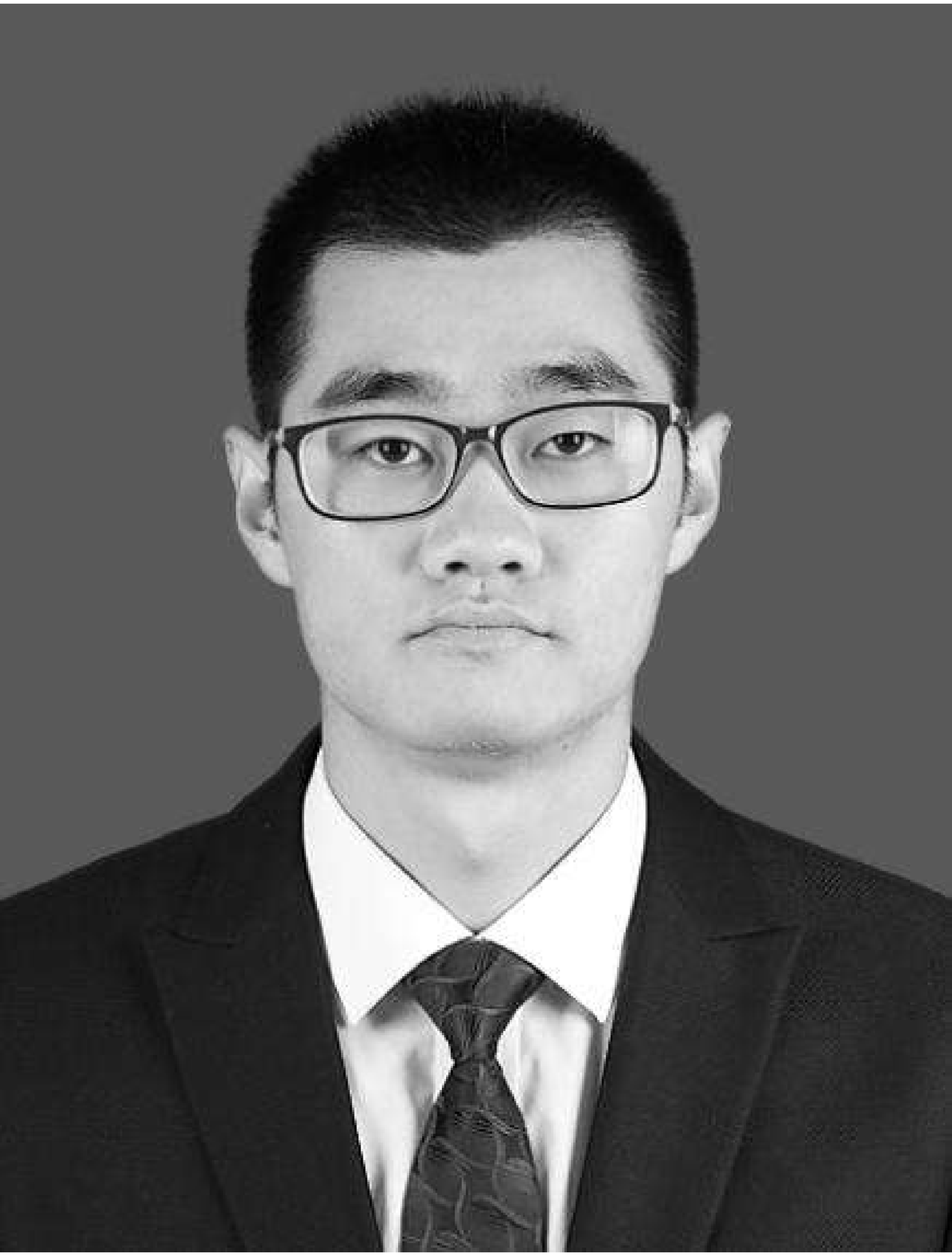}}]{Ning Li} received the B.S degree from the School of Automation, Northwestern Polytechnical University, Xi’an, in 2016, where he is currently pursuing the Ph.D. degree. 

His current research interests include polarization imaging and computer vision, especially visual perception based on multiband and polarization information.
\end{IEEEbiography}

\begin{IEEEbiography}[{\includegraphics[width=1in,height=1.25in,clip,keepaspectratio]{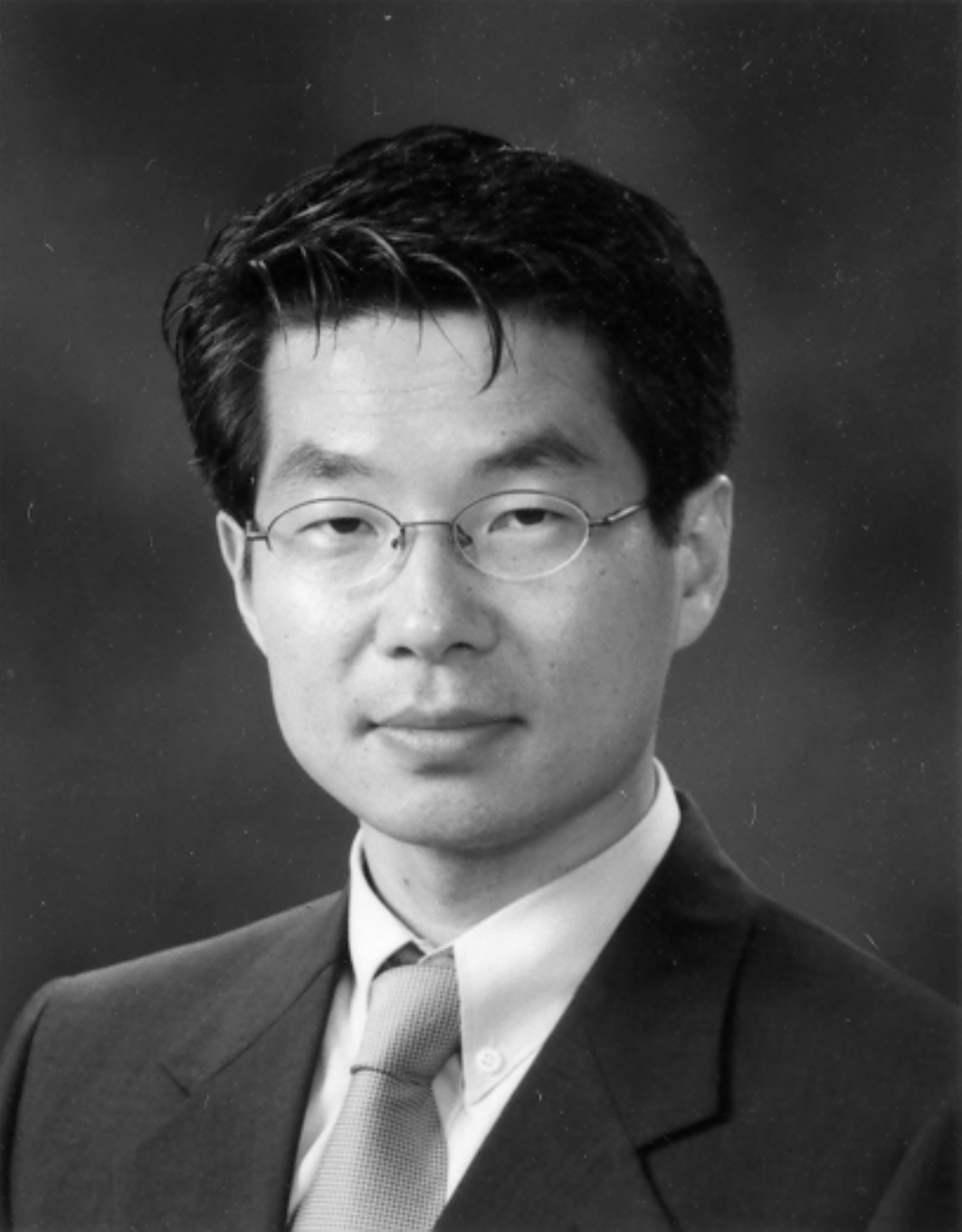}}]{Seong G. Kong} (Senior Member, IEEE) received
	the B.S. and M.S. degrees in electrical engineering from Seoul National University, Seoul, South
	Korea, in 1982 and 1987, respectively, and the Ph.D.
	degree in electrical engineering from the University
	of Southern California, Los Angeles, CA, USA,
	in 1991.
	
	He was an Associate Professor of electrical and
	computer engineering with the University of Tennessee, Knoxville, TN, USA, and Temple University,
	Philadelphia, PA, USA. He was also the Chair with
	the Department of Electrical Engineering, Soongsil University, Seoul, and
	the Graduate Program Director with the Electrical and Computer Engineering
	Department, Temple University. He is a Professor of computer engineering and
	the Director of strategic planning with Sejong University, Seoul. His research
	interests include image processing, computer vision, hyperspectral imaging,
	and intelligent systems.
	
	Dr. Kong was a recipient of the Honorable Mention Paper Award from the
	American Society of Agricultural and Biological Engineers in 2005 and the
	Most Cited Paper Award from Computer Vision and Image Understanding
	in 2007 and 2008. He was an Associate Editor of the IEEE TRANSACTIONS
	ON NEURAL NETWORKS and the Guest Editor of the Journal of Sensors.
\end{IEEEbiography}

\bibliographystyle{IEEEtran}
\bibliography{IEEEabrv,IEEEexample}

\end{document}